# Forecasting Cardiology Admissions from Catheterization Laboratory

Abstract ID: 570456

Avishek Choudhury
School of Systems and Enterprises
Stevens Institute of Technology
Hoboken, NJ

Sunanda Perumalla
Texas Tech University
Lubbock,TX

Dr. Christopher M. Greene
Binghamton University
Binghamton, NY

## Abstract

Emergent and unscheduled cardiology admissions from cardiac catheterization laboratory add complexity to the management of Cardiology and inpatient departments. In this article, the authors sought to study the behavior of cardiology admissions from Catheterization laboratory using time series models. This research involves retrospective cardiology admission data from March 1, 2012, to November 3, 2016, retrieved from a hospital in Iowa. Autoregressive integrated moving average (ARIMA), Holt's method, mean method, naïve method, seasonal naïve, exponential smoothing, and drift method were implemented to forecast weekly cardiology admissions from Catheterization laboratory. ARIMA (2,0,2) (1,1,1) was selected as the best fit model with the minimum sum of error, Akaike information criterion, and Schwartz Bayesian criterion. The model failed to reject the null hypothesis of stationarity; it lacked the evidence of independence and rejected the null hypothesis of normality. The implication of this study will not only improve catheterization laboratory staff scheduling and advocate efficient use of imaging equipment and inpatient telemetry beds, but will also equip management to proactively address inpatient overcrowding, plan for physical capacity expansion, and so forth.

## 1. Introduction

Catheterization laboratories are key capital and labor-intensive subdivisions within a hospital. These subdivisions are fiscally crucial for a hospital. Due to the mounting predominance of cardiovascular disorders and burgeoning types of catheter instances, the demand for catheterization procedures is at its peak [1]. The number of cardiac catheterizations executed in the United States has escalated significantly over the last 30 years. Catheterization laboratory infrastructure has improved, and during 2007 about 85% of all United States hospitals delivered cardiac catheterization assistance [2]. Although the volume of Catheterization laboratories has decreased in recent years, the assortment of catheterization procedures has burgeoned to encompass both diagnostic and therapeutic procedures [2]. A diverse amalgam of procedure types and complex patient health conditions makes it arduous to predict patients' post-procedure care needs. Thus, before catheterization, it is unknown whether a patient's post-procedure health situation will necessitate an inpatient overnight stay [3]. Uncertainty concerning cardiology admissions from catheterization laboratory defies efficient inpatient bed management and laboratory resource utilization. Cardiology hospital units contain specialized monitoring equipment (telemetry beds) designed for recovering catheterization patients, in addition to cardiac patients admitted from the emergency department (ED) and external locations (direct admissions). To proficiently manage patient flow, providers must project the number of admissions from each of these sources daily. The Catheterization laboratory is an enormous source of cardiac inpatient admissions, and due to the high uncertainty of admission numbers from the Catheterization laboratory, daily as well as weekly overall admission projections often focus on this source [2]. In several hospitals, providers depend on their intuition and experience to envisage the demand for beds (the number of admissions) for a given day or week after assessing the catheterization schedule of the same day or week [4]. Cardiology admission from catheterization laboratory prognostication is a precarious component of hospital resource management that could be supported by time series forecasting.

The objective of this study is to develop and prospectively validate a decision support application to predict cardiology admissions for adults from a catheterization laboratory. This study implements the ARIMA to forecast admissions to catheterization laboratory. ARIMA is one of the most effective techniques of time series forecasting [5]. This method has been widely used in various fields such as forecasting energy consumption [6], stock prices [7], population growth [8], and many more. However, the application of the time series technique of ARIMA for forecasting cardiology admissions from catheterization laboratory is untapped.

## 2. Methodology

Our team retrospectively extracted 244 weeks of cardiology admission data from March 1, 2012, through November 3, 2016, from a hospital in Iowa [9]. The time series analysis was conducted on the first 200 weeks (March 1, 2012, through December 31, 2015), and the forecasted values were matched against the actual patient admission (March 1, 2012, through November 3, 2016). During the process, the Kwiatkowski–Phillips–Schmidt–Shin (KPSS) test, Phillips-Perron unit root test, and augmented Dickey-Fuller test were performed to determine whether the time series was stationary around a mean or linear trend or is non-stationary owing to a unit root, while white noise was tested using the Box–Ljung test. Moreover, we employed the Shapiro-Wilk test, Cramer-von Mises test, Kolmogorov-Smirnov test, Pearson chi-squared test, Shapiro-Francia test, and the Anderson–Darling test for normality. This was then followed by plotting of the autocorrelation function and partial autocorrelation function to determine ARIMA (p, d, q), where p is the order of autoregression (AR), d is the lagged difference between the current and previous values, and q denotes the order of the moving average (MA). The maximum values of *p* and *q* were set as 24, and the maximum number of non-seasonal difference (d) was set as 4. ARIMA models are, in theory, a general class of models for forecasting a time series which can be crafted to be "stationary" by differencing, conceivably in conjunction with non-linear transformations such as lagging or deflating.

A time series is stationary if its statistical properties are all constant over time. A stationary series has no trend, its variations around its mean have a constant amplitude, and it wriggles steadily; in other words, its short-term random time patterns always look the same in a statistical sense. The latter state indicates that its autocorrelations (correlations with its prior deviations from the mean) stay constant over time. A random variable of this form can be perceived as a combination of signal and noise, and the signal could be an array of fast or slow mean reversion, or sinusoidal oscillation, or rapid alternation in sign; and it could have a seasonal component.

An ARIMA model can be viewed as a "filter" that separates the signal from the noise, and the signal is then extrapolated into the future to obtain forecasts. Let y denote the $d^{th}$ difference of Y, which means: if d=0: $y_t = Y_t$; if d=1: $y_t = Y_t - Y_{t-1}$; if d=2: $y_t = (Y_t - Y_{t-1}) - (Y_{t-1} - Y_{t-2}) = Y_t - 2Y_{t-1} + Y_{t-2}$; in terms of y, the general forecasting equation is (eq. 1):

$$\hat{y}_t = \mu + \phi_1 y_{t-1} + \cdots + \phi_p y_{t-p} - \theta_1 e_{t-1} - \cdots - \theta_q e_{t-q} \qquad (1)$$

Here the moving average parameters (θ's) are defined so that their signs are negative in the equation, following the convention introduced by Box and Jenkins. However, the R programming language defines them so that they have plus signs instead.

Additionally, Naïve, seasonal Naïve, mean, exponential smoothing, drift, and Holt's methods were also implemented and compared with ARIMA. The fitted model with minimum Akaike information criterion (AIC) and Schwartz Bayesian criterion (BIC) was selected as the optimal forecasting model [10]. The AIC is an estimator of the relative quality of statistical models for a given set of data. Given a compendium of models for the data, AIC evaluates the quality of each model, relative to the other models. The AIC is established on information theory. When a statistical model is used to epitomize the data generating process, the representation incorporates information loss. AIC estimates the relative information lost by a given model: the less information a model squanders, the superior the quality of that model. In estimating the information lost, AIC deals with the trade-off between the goodness of fit of the model and the austerity of the model. BIC, also known as Schwarz information criterion (also SIC, SBC, SBIC) is a benchmark for model selection among a finite set of models; the model with the lowest BIC is preferred. It is based, in part, on the likelihood function and it is closely related to the AIC. The accuracy of the forecast was then measured based on the sum of errors.

The Ljung-Box statistic, also known as modified Box-Pierce statistic, is a function of accumulated sample auto-correlations ($r_j$) up to any specified time lag (m).

$$Q(m) = n(n+2) \sum_{j=1}^{m} \frac{r^2}{n-j} \quad (2)$$

where n is the number of usable data points differencing operation.

The Unit root tests are assessments for stationarity in a time series. A time series has stationarity if a shift in time does not instigate an alteration in the shape of the distribution. It is a stochastic trend in a time series. If a time series has a unit root, it shows a systematic pattern that is typically unpredictable. Unit root test implicitly assumes that the time series $[y_t]_{t=1}^{T}$ to be tested can be written as (eq.3),

$$y_t = D_t + z_t + \epsilon_t \quad (3)$$

where, $D_t$ is the deterministic component (trend, seasonal component, etcetera); $z_t$ is the stochastic component; $\epsilon_t$ is the stationarity error process. Common unit root tests include KPSS test and Augmented Dickey-Fuller test.

Augmented Dickey-Fuller (ADF) test is an augmented version of the dickey-fuller test for a more extensive and elaborate time series models. ADF tests the null hypothesis that a unit root is present in a time series sample. The alternative hypothesis is stationarity or trend-stationarity. ADF is applied to the model (eq.4),

$$\Delta_{y_t} = \alpha + \beta t + \gamma y_{t-1} + \delta_1 \Delta y_{t-1} + \cdots + \delta_{P-1} \Delta y_{t-P+1} + \varepsilon_t \quad (4)$$

where $\alpha$ is constant, $\beta$ is the coefficient on a time trend, and $p$ is the lag order of the autoregressive process. The instinct behind this test is if the series is integrated, then the lagged level of the series $y_{t-1}$ will provide no relevant information in predicting the change in $y_t$ besides the one obtained in the lagged changes $\Delta y_{t-k}$.

KPSS analyzes whether a time series is stationary around a mean or linear trend or is non-stationary due to a unit root. The null hypothesis of KPSS is that an observable time series is stationary around a deterministic trend and alternative. This is contrary to most of the commonly used unit root tests. The KPSS test is based on linear regression, and it breaks up a series into three parts (eq.5).

$$x_t = r_t + \beta_t + \varepsilon_t \quad (5)$$

where $r_t$ is a random walk, $\beta_t$ is a deterministic trend, $\varepsilon_t$ is a stationary error.

## 3. Results
### 3.1. Temporal Analysis
The weekly data containing cardiology admissions from catheterization laboratory from March 1, 2012, through November 3, 2016, shows an increasing trend over the past years and a cyclic seasonal effect. The following figure 1 depicts the seasonal plot for the same. The figure shows the non-polar view of the seasonal plot.

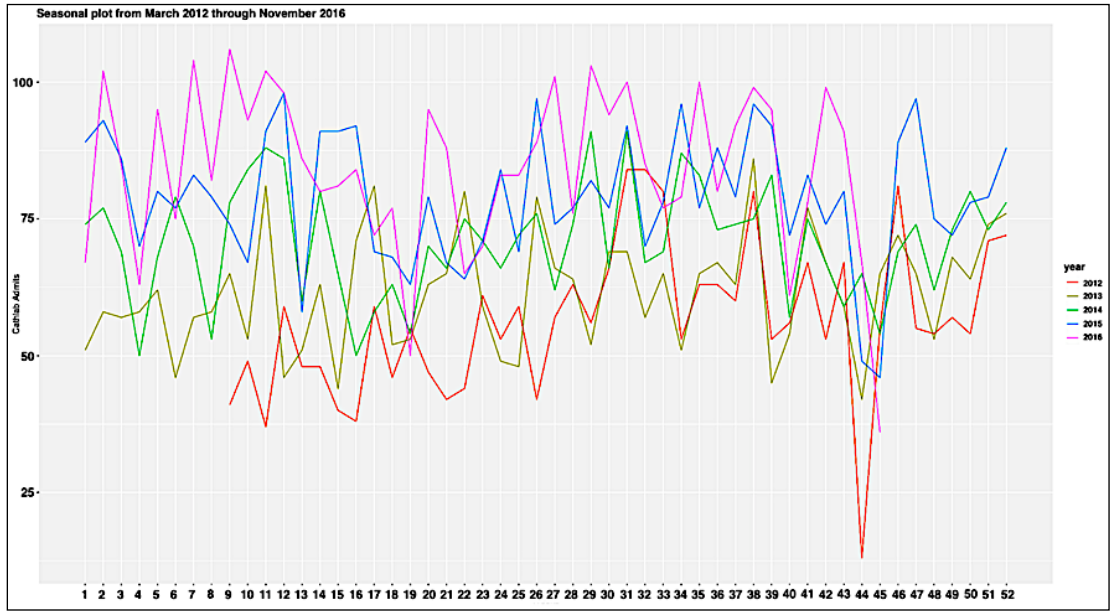

Figure 1: X-axis shows the number of weeks ranging from 1 through 52 and Y-axis indicates the number of admissions from the catheterization laboratory. The figure shows the seasonal plot of weekly admissions from catheterization laboratory from March 1, 2012, through November 3, 2016. The plot shows the fluctuating weekly trend over 52 weeks and cyclic seasonality.

### 3.2. Comparative analysis of forecasting methods

The best fit model ARIMA (2,0,2) (1,1,1) with AIC of 1528, BIC of 1553, and minimum error of 27.90 was selected. Table 1 compares the sum of the errors of Naïve, seasonal Naïve (SN), mean, exponential smoothing (SES), drift, and Holt's method and ARIMA. Table 2 describes the selected ARIMA model.

Table 1: Comparative analysis

| MODELS | ARIMA | HOLT | SES | MEAN | DRIFT | NAIVE | SN |
|---|---|---|---|---|---|---|---|
| SUM OF ERROR | 27.9 | 34.0 | 37.7 | 44.6 | 45.4 | 45.7 | 60.7 |

Table 2: ARIMA (2,0,2) (1,1,1)

| | ar1 | ar2 | Ma1 | Mar2 | sar2 | sma1 | drift |
|---|---|---|---|---|---|---|---|
| Coefficients | -0.03 | 0.55 | 0.09 | -0.58 | -0.11 | -0.42 | 0.14 |
| s.e | 0.87 | 0.65 | 0.87 | 0.69 | 0.21 | 0.23 | 0.01 |

### 3.3. Model Evaluation

All the normality test of residuals yielded a p-value of less than *0.05*. Thus, the tests reject the null hypothesis of normality implying the residuals are not normally distributed. The values of all the normality tests are listed in the following table 3.

Table 3: Normality tests

| Normality Test | p-value (0.05) | Null Hypothesis of Normality |
|---|---|---|
| Anderson-Darling | 1.02 e$^{-08}$ | Rejects null hypothesis |
| Shapiro-Wilk | 8.38 e$^{-05}$ | Rejects null hypothesis |
| Cramer-Von Mises | 2.55 e$^{-08}$ | Rejects null hypothesis |
| Kolmogorov-Smirnov | 7.13 e$^{-12}$ | Rejects null hypothesis |
| Pearson Chi-Square | 2.20 e$^{-16}$ | Rejects null hypothesis |
| Shapiro-Francia | 0.0001 | Rejects null hypothesis |

The Box–Ljung test on residuals gave a p-value of *0.73* (*p-value > 0.05*), which implies that there is no significant autocorrelation. In other words, there is a lack of evidence of independence. Figure 2 below shows the autocorrelation and partial autocorrelation plot.

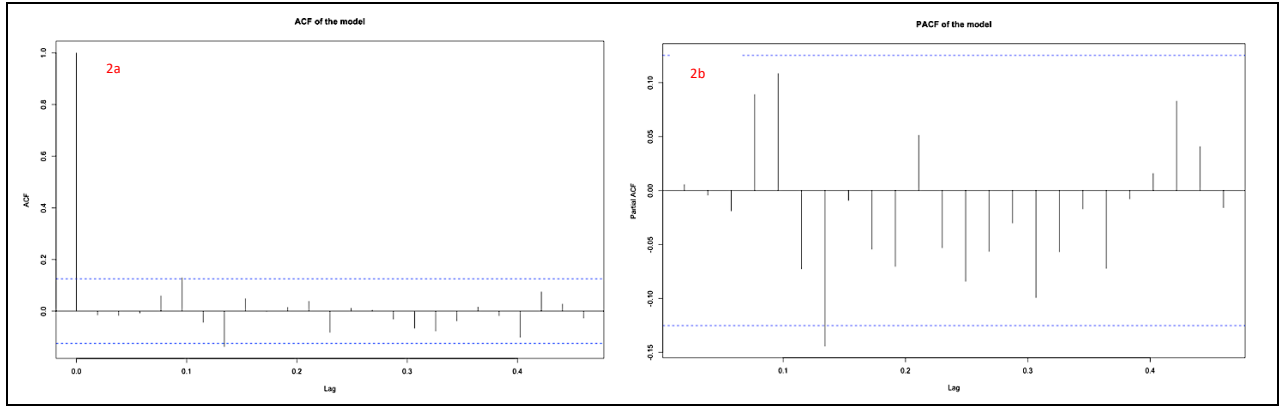

Figure 2: (a) shows the ACF plot, (b) shows the PACF plot with one point exceeding the lower limit at *0.14*

The augmented Dickey-Fuller test, and Phillips-Perron unit root test yielded a p-value of *0.01* each (*p-value < 0.05*); thus, the test rejects the null hypothesis of non-stationarity. Additionally, Kwiatkowski-Phillips Schmidt Shin test with a p-value of *0.1* fails to reject the null hypothesis of stationarity.

### 3.4. Forecast Accuracy

As discussed above, the best weekly forecast accuracy was exhibited by ARIMA (2,0,2) (1,1,1). We forecasted *52* weeks (December 3, 2016, through December 2, 2017) ahead of the actual dataset as shown in figure 3 below. The following figure 4 shows a visual comparison of the forecasted and actual cardiology admissions from catheterization laboratory visits.

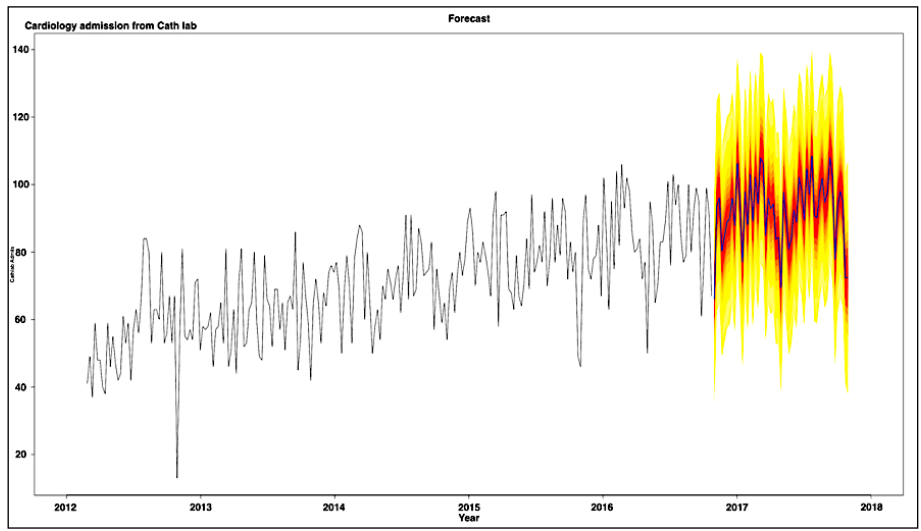

Figure 3: X-axis represents the years, and Y-axis indicates the number of cardiology admission from the catheterization laboratory. The figure shows the forecasted cardiology admission from catheterization laboratory (*December 3, 2016, through December 2, 2017*). The color shades in the figure show the different confidence intervals starting from 10% through 99%.

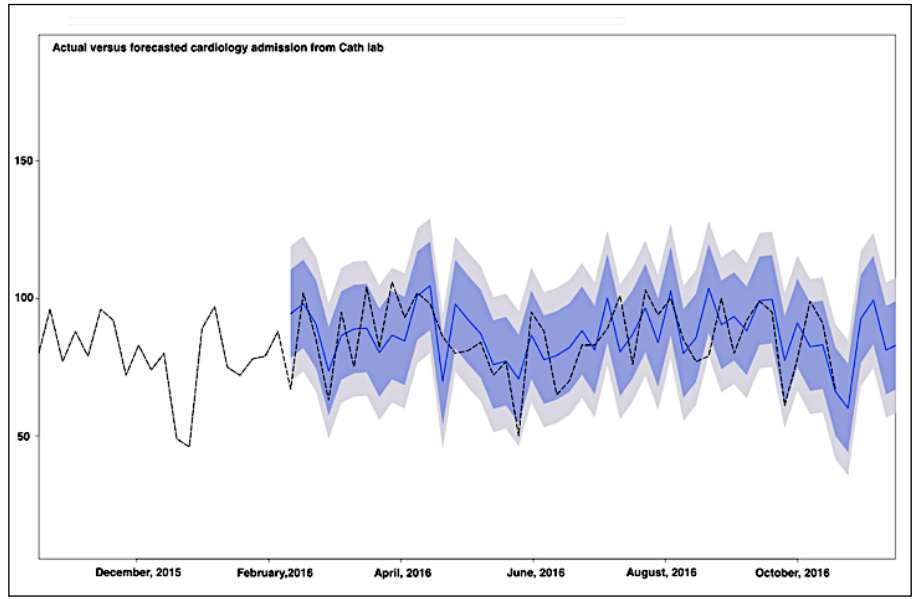

Figure 4: X-axis shows the timeline *(January 30, 2016, through December 3, 2016)* and Y-axis represents the cardiology admission from catheterization laboratory. The figure compared the actual admission (black line) and forecasted admission (blue line) to cardiology from catheterization laboratory at 80% and 95% upper and lower confidence intervals.

## 4. Discussion

The analyses performed in this study focused on a time series forecasting method using ARIMA in RStudio. The analyses were implemented for forecasting weekly cardiology admissions from catheterization laboratory, and a best fit ARIMA model was proposed. Based on AIC and BIC, the best fit model was found to be ARIMA (2,0,2) (1,1,1). ARIMA outperformed Holt's method, exponential smoothing, mean method, drift method, Naïve method, and seasonal Naïve for this dataset. The model proposed in this study tests for normality, stationarity, and autocorrelation. Several other studies have employed time series forecasting methodology with high accuracy to forecast hourly, weekly, monthly and yearly arrivals of patients visiting an emergency department. This is the first study concerning weekly forecasting of cardiology patient admission from the catheterization laboratory. The proposed model can help clinical staffs and management proactively deal with fluctuating patient admission rate. Predicting weekly visits provides a useful overall projection of total admissions for a month and also projects an acceptable annual telemetry bed demand in cardiology department or floor. All these proactive measures can help minimize overcrowding and advocate appropriate resource allocation.

## 5. Conclusion

ARIMA (2,0,2) (1,1,1) was the best fit model to forecast cardiology admission from catheterization laboratory and can be utilized as a decision support system in the healthcare industry. For future research, time series forecasting method such as neural networks, fuzzy logic, and TABTS shall be implemented and compared against ARIMA. Consequently, hourly forecasting cardiology admission from catheterization laboratory can be used to manage staff schedules, bed allotment, and optimize inpatient flow.

### Data statement
Anonymized weekly cardiology admission from Catheterization laboratory data (March 1, 2012, through November 3, 2016) is available at *DOI:10.17632/wgz36h39wt.2*.